\documentclass[10pt,journal]{IEEEtran}
\usepackage{graphicx}
\usepackage{amsfonts}
\usepackage{amsmath}
\usepackage{amssymb}
\usepackage{color}
\usepackage{graphicx,color,overpic}
\usepackage{psfrag}
\usepackage{amssymb}
\usepackage{times}
\usepackage{latexsym}
\usepackage{bm}
\usepackage{cases}
\usepackage{array}
\usepackage{fancyhdr}
\usepackage{setspace}
\usepackage{subfigure}
\usepackage{url}
\usepackage{algorithm}
\usepackage{algorithmic}
\usepackage{multirow}
\usepackage{epstopdf}

\usepackage{epsfig}
\usepackage{fancybox}
\usepackage{textcomp}

\IEEEoverridecommandlockouts
\IEEEpubid{\makebox[\columnwidth]{978-1-4799-5674-6/14/\$31.00~\copyright~2014 IEEE \hfill}
\hspace{\columnsep}\makebox[\columnwidth]{ }}

\begin{document}

\title{Reconfigurable Intelligent Surface Assisted Multiuser MISO Systems Exploiting Deep Reinforcement Learning}

\author{Chongwen Huang,~\IEEEmembership{Member,~IEEE}, Ronghong Mo and  Chau Yuen,~\IEEEmembership{Senior Member,~IEEE}

\thanks{The work of C. Yuen and C. Huang was supported by $A^*$STAR under its RIE2020 Advanced Manufacturing and Engineering (AME) Industry Alignment Fund 每 Pre Positioning (IAF-PP) (Grant No. A19D6a0053). (\emph{Corresponding author: Ronghong Mo}.)

C.~Huang, R. Mo and C. Yuen are with the Singapore University of Technology and Design, 487372 Singapore (emails: chongwen\_huang@sutd.edu.sg, yuenchau@sutd.edu.sg).} }

\maketitle

\begin{abstract}
Recently, the reconfigurable intelligent surface (RIS), benefited from the breakthrough on the fabrication of programmable meta-material, has been speculated as one of the key enabling technologies for the future six generation (6G) wireless communication systems scaled up beyond massive multiple input multiple output (Massive-MIMO) technology to achieve smart radio environments. Employed as reflecting arrays, RIS is able to assist MIMO transmissions without the need of radio frequency chains resulting in considerable reduction in power consumption. In this paper, we investigate the joint design of transmit beamforming matrix at the base station and the phase shift matrix at the RIS, by leveraging recent advances in deep reinforcement learning (DRL). We first develop a DRL based algorithm, in which the joint design is obtained through trial-and-error interactions with the environment by observing predefined rewards, in the context of continuous state and action. Unlike the most reported works utilizing the alternating optimization techniques to alternatively obtain the transmit beamforming and phase shifts, the proposed DRL based algorithm obtains the joint design simultaneously as the output of the DRL neural network. Simulation results show that the proposed algorithm is not only able to learn from the environment and gradually improve its behavior, but also obtains the comparable performance compared with two state-of-the-art benchmarks. It is also observed that, appropriate neural network parameter settings will improve significantly the performance and convergence rate of the proposed algorithm.
\end{abstract}


\vspace{-0.25cm}
\section{Introduction}
Recent years have witnessed the successful deployment of massive multiple input multiple output (Massive-MIMO) in the fifth generation (5G)  wireless communication systems, as a promising approach to support massive number of users at high data rate, low latency and secure transmission simultaneously and efficiently \cite{Ref1}-\cite{Ref3}. However, implementing a Massive-MIMO base station (BS) is challenging, as high hardware cost, constrained physical size, and increased power consumption scaling up the conventional MIMO systems by many orders of magnitude, arise when the conventional large-scale antenna array is used at the BS.


On another hand, the reconfigurable intelligent surface (RIS), benefited from the breakthrough on the fabrication of programmable meta-material, has been speculated as one of the key enabling technologies for the future six generation (6G) wireless communication systems scaled up beyond Massive-MIMO to achieve smart radio environment \cite{chongwentwc2019}-\cite{Ref7}. The meta-material based RIS makes possible wideband antennas with compact size, such that large scale antennas can be easily deployed at both ends of the user devices and BS, to achieve Massive-MIMO gains but with significant reduction in power consumption. With the help of varactor diode or other micro electrical mechanical systems (MEMS) technology, Electromagnetic (EM) properties of the RIS are fully defined by its micro-structure, and can be programmed to vary the phase, amplitude, frequency and even orbital angular momentum of an EM wave, effectively modulating a radio signal without a mixer and radio frequency (RF) chain.

The RIS can be deployed as reconfigurable transmitters, receivers and passive reflecting arrays. Being reflecting arrays, the RIS is usually placed in between the BS and single antenna receivers, and consists of a vast number of nearly passive, low-cost and low energy consuming reflecting elements, each of which introduces a certain phase shift to the signals impinging on it. By reconfiguring the phase shifts of elements of RIS, the reflected signals can be added constructively at the desired receiver to enhance the received signal power or destructively at non-intended receivers to reduce the co-channel interference. Due to the low power consumption, the reflecting RIS can be fabricated in very compact size with light weight, leading to easy installation of RIS in building facades, ceilings, moving trains, lamp poles, road signs, etc., as well as ready integration into existing communication systems with minor modifications on hardware \cite{Ref7}-\cite{Ref10}.

Note that, passive reflecting surfaces have been used in radar systems for many years. However, the phase shifts of passive radars cannot be adjusted once fabricated, and the signal propagation cannot be programmed through controlling the phase shifts of antenna elements. The reflecting RIS also differs from relaying systems, in that the RIS reflecting array only alters the signal propagation by reconfiguring the constituent meta-atoms of meta-surfaces of RISs without RF chains and additional thermal noise added during reflections, whereas the latter requires active RF components for signal reception and emission. Consequently, the beamforming design in relay nodes is classified as active, while it is passive in reflecting RIS assisted systems.

\subsection{Prior Works}
Although RIS has gained considerable attentions in recent years, the most of reported works are primarily focused on implementing hardware testbeds, e.g., reflect-arrays and meta-surfaces, and on realizing point-to-point experimental tests \cite{Ref6}-\cite{Ref7}. More recently, there are some works attempting to investigate optimizing the performance of RIS-assisted MIMO systems. The optimal receiver and matched filter (MF) were investigated for uplink RIS assisted MIMO systems in \cite{Ref5}, where the RIS is deployed as a MIMO receiver. An index modulation (IM) scheme exploiting the programmable nature of the RIS was proposed in \cite{Ref9}, where it was shown that the RIS-based IM enables high data rates with remarkably low error rates.

When RISs are utilized as reflecting arrays, the error performance achieved by a reflecting RIS assisted single antenna transmitter/receiver system was derived in \cite{Ref10}. A joint design of local optimal transmit beamforming at the BS and the phase shifts at reflecting RIS with discrete entries was proposed in \cite{Ref10a} for reflecting RIS assisted single-user multiple input single output (MISO) systems, by solving the transmit power minimization problem utilizing an alternating optimization technique. The received signal power maximization problem for MISO systems with reflecting RIS was formulated and studied in \cite{Ref10b} through the design of transmit beamforming and phase shifts employing efficient fixed point iteration and manifold optimization techniques. The authors in \cite{Ref10c} derived a closed-form expression for the phase shifts for reflecting RIS assisted MISO systems when only the statistical channel state information (CSI) is available. Compressive sensing based channel estimation was studied in \cite{Ref10d} for reflecting RIS assisted MISO systems with single antenna transmitter/receiver. Deep learning based algorithm was  proposed to obtain phase shifts. In \cite{Ref10f}-\cite{Ref10g}, the transmit beamforming and the phase shifts were designed to maximize the secrecy rate for reflecting RIS assisted MIMO systems with only one legitimate receiver and one eavesdropper, employing various optimization techniques.

All above mentioned works focus on single user MISO systems. As multiple users and massive access are concerned, the transmit beamforming and the phase shift were studied in \cite{Ref10e}-\cite{Ref11}, by solving the sum rate/energy efficiency maximization problem, assuming a zero-forcing (ZF) based algorithm employed at the BS, whereas the stochastic gradient descent (SGD) search and sequential fractional programming are utilized to obtain the phase shifter. In \cite{Ref11a}, through minimizing the total transmit power while guaranteeing each user's signal-to-interference-plus-noise ratio (SINR) constraint, the transmit beamforming and phase shifts were obtained by utilizing semi-definite relaxation and alternating optimization techniques. In \cite{Ref11b}, the fractional programming method was used to find the transmit beamforming matrix, and three efficient algorithms were developed to optimize the phase shifts. In \cite{Ref12}, the large system analysis was exploited to derive the closed-form expression of the minimum SINR when only spatial correlation matrices of the RIS elements are available. Then, authors targeted at maximizing the minimum SINR by optimizing the phase shifts based on the derived expression. In \cite{Ref13b}, the weighted sum rate of all users in multi-cell MIMO settings were investigated, through jointly optimizing the transmit beamforming and the phase shifts subject to each BS's power constraint and unit modulus constraint.

Recently, the model-free artificial intelligence (AI) has emerged as an extraordinarily remarkable technology to address explosive mass data, mathematically intractable non-linear non-convex problems and high-computation issues \cite{Ref14}-\cite{Ref14c}. Overwhelming interests in applying AI to the design and optimization of wireless communication systems have been witnessed recently, and it is a consensus that AI will be at the heart of future wireless communication systems (e.g. 6G and beyond) \cite{Ref15}-\cite{Ref16g}. The AI technology is most appealing to large scale MIMO systems with massive number of array elements, where optimization problems become non-trivial due to extremely large dimension optimization involved. Particularly, deep learning (DL) has been used to obtain the beamforming matrix for MIMO systems by building a mapping relations between channel information and the precoding design \cite{Ref16a}-\cite{Ref16d}. Actually, DL based approaches are able to significantly reduce the complexity and computation time utilizing the offline prediction, but often require an exhaustive sample library for online training. Meanwhile, the deep reinforcement learning (DRL) technique which embraces the advantage of DL in neural network training as well as improves the learning speed and the performance of reinforcement learning (RL) algorithms, has also been adopted in designing wireless communcation systems \cite{Ref14b}, \cite{Ref15a}, \cite{Ref16e}-\cite{Ref16g}.

DRL is particularly beneficial to wireless communication systems where radio channels vary over time. DRL is able to allow wireless communication systems to learn and build knowledge about the radio channels without knowing the channel model and mobility pattern, leading to efficient algorithm designs by observing the rewards from the environment and find out solutions of sophisticated optimization problems. In \cite{Ref16e}, the hybrid beamforming matrices at the BS were obtained by applying DRL where the sum rate and the elements of the beamforming matrices are denoted as states and actions. In \cite{Ref16g}, the cell vectorization problem is casted as the optimal beamforming matrix selection to optimize network coverage utilizing DRL to track the user distribution pattern. In \cite{Ref16f}, the joint design of beamforming, power control, and interference coordination were formulated as an non-convex optimization problem to maximize the SINR solved by DRL.
\subsection{Contributions}
In this paper, we investigate the joint design of transmit beamforming at the BS and phase shifts at the reflecting RIS to maximize the sum rate of multiuser downlink MISO systems utilizing DRL, assuming that direct transmissions between the BS and the users are totally blocked. This optimization problem is non-convex due to the multiuser interference, and the optimal solution is unknown. We develop a DRL based algorithm to find the feasible solution, without using sophisticate mathematical formulations and numerical optimization techniques. Specifically, we use policy-based deep deterministic policy gradient (DDPG) derived from Markov decision process to address continuous beamforming matrix and phase shifts \cite{Ref17}. The main contributions of this paper are summarized as follows:

$\bullet$ We propose a new joint design of transmit beamforming and phase shifts based on the recent advance in DRL technique. This paper is a very early attempt to formulate a framework that incorporates the DRL technique into optimal designs for reflecting RIS assisted MIMO systems to address large-dimension optimization problems.

$\bullet$ The proposed DRL based algorithm has a very standard formulation and low complexity in implementation, without knowledge of explicit model of wireless environment and specific mathematical formulations. Such that it is very easy to be scaled to various system settings. Moreover, in contrast to DL based algorithms which rely on sample labels obtained from mathematically formulated algorithms, DRL based algorithms are able to learn the knowledge about the environment and adapt to the environment.

$\bullet$ Unlike reported works which utilize alternating optimization techniques to alternatively obtain the transmit beamforming and phase shifter, the proposed algorithm jointly obtain the transmit beamforming matrix and the phase shifts, as one of the outputs of the DRL algorithm. Specifically, the sum rate is utilized as the instant rewards to train the DRL based algorithm. The transmit beamforming matrix and the phase shifts are jointly obtained by gradually maximizing the sum rate through observing the reward and iteratively adjusting the parameters of the proposed DRL algorithm accordingly. Since the transmit beamforming matrix and the phase shifts are continuous, we resort to DDPG to develop our algorithm, in contrast to designs addressing the discrete action space.

Simulations show that the proposed algorithm is able to learn from the environment through observing the instant rewards and improve its behavior step by step to obtain the optimal transmit beamforming matrix and phase shifts. It is also observed that, appropriate neural network parameter settings will increase significantly the performance and convergence rate of the proposed algorithm.

The rest of the paper is organized as follows. The system model will be described in Section II. The DRL based algorithm for joint design of transmit beamforming  and phase shifts is presented in Section III. Simulation results are provided in Section V to verify the performance of the proposed algorithms, whereas conclusions are presented in Section VI.

The notations used in this paper are listed as follows. $\mathcal{E}$ denotes the statistical expectation. For any general matrix $\mathbf{H}$, $\mathbf{H}(i,j)$ denotes the entry at the $i^{th}$ row and the $j^{th}$ column. $\mathbf{H}^T$, and $\mathbf{H}^{\mathcal{H}}$ represent the transpose and conjugate transpose of matrix $\mathbf{H}$, respectively. $\mathbf{H}^{(t)}$ is the value of $\mathbf{H}$ at time $t$. $\mathbf{h}_k$ is the $k^{th}$ column vector of $\mathbf{H}$. $Tr \{ \}$ is the trace of the enclosed item. For any column vector $\mathbf{h}$ (all vectors in this paper are column vectors), $\mathbf{h}(i)$ is the $i^{th}$ entry, while $\mathbf{h}_{k,n}$ is the $n^{th}$ channel vector for the $k^{th}$ user. $||\mathbf{h}||$ denotes the magnitude of the vector. $|x|$ denotes the absolute value of a complex number ${x}$, and $Re{(x)}$ and $Im{(x)}$ denote its real part and imaginary part, respectively.

\vspace{-0.15cm}
\section{System Model and Problem Formulation}
\begin{figure}
 \begin{center}
  \includegraphics[width=9.3cm]{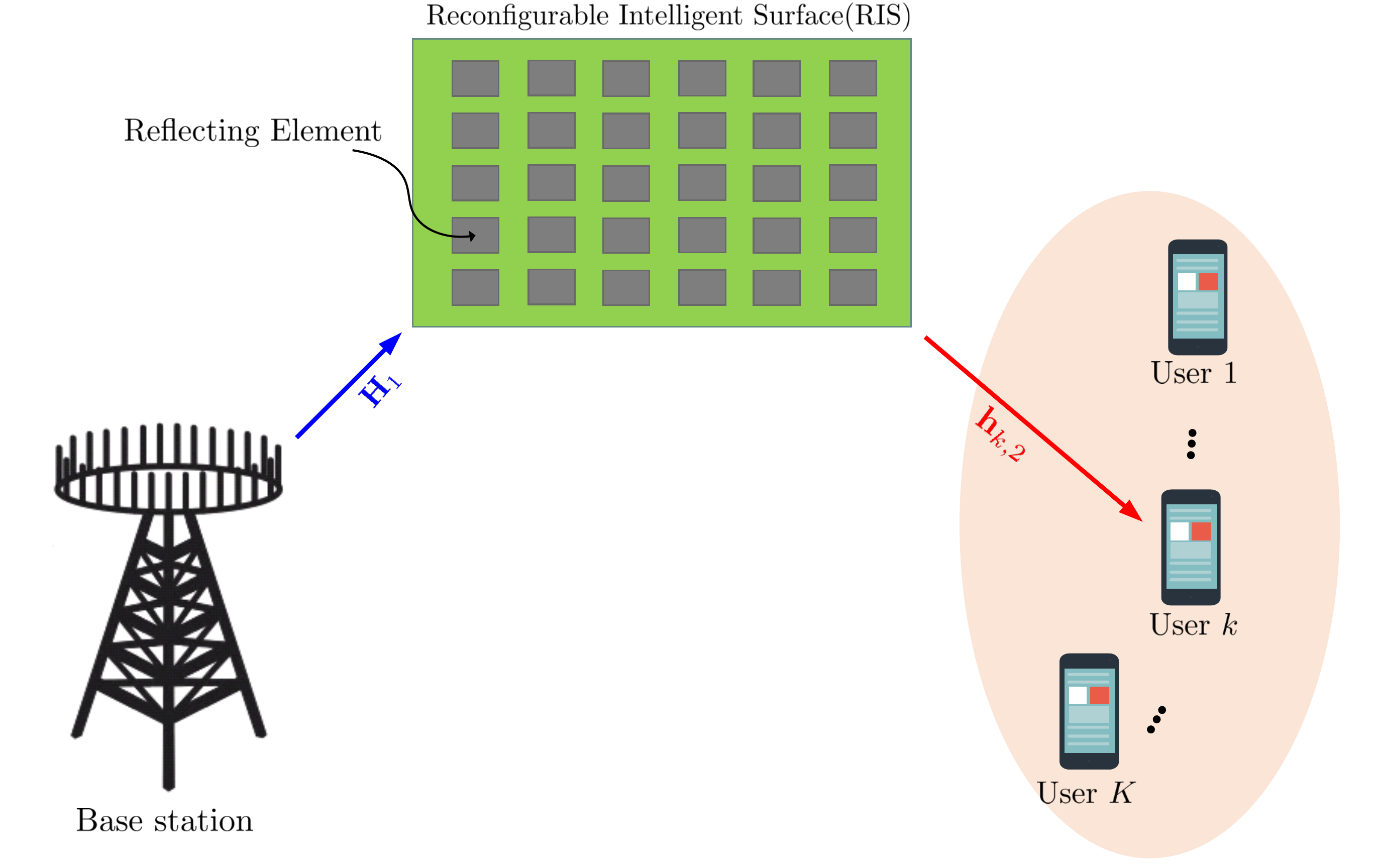}
  \caption{The considered RIS-based multi-user MISO system comprised of a $M$-antenna BS simultaneously serving in the downlink $K$ single-antenna users. RIS is equipped with $N$ reflecting elements and one micro-controller, which is  attached to a building's facade, and the transmit signal propagates to the users via the RIS's assistance.}
  \label{fig:SM}
 \end{center}\vspace{-6mm}
\end{figure}

We consider a MISO system comprised of a BS, one reflecting RIS and multiple users, as shown in Fig. \ref{fig:SM}. The BS has $M$ antennas and communicates with $K$ users where $M \geq K$ single antenna users. The reflecting RIS is equipped with $N$ reflecting elements and one micro-controller. A number of $K$ data streams are transmitted simultaneously from the $M$ antennas of the BS. Each data stream is targeted at one of the $K$ users. The signals are first arrived at the reflecting RIS and then are reflected by the RIS. The direct signal transmissions between the BS and the users are assumed to be negligible. This is reasonable since in practical the reflecting RIS is generally deployed to overcome the situations where severe signal blockage happens in-between the BS and the users. The RIS functions as a reflecting array, equivalent to introducing phase shifts to impinging signals. Being intelligent surface, the reflecting RIS could be intelligently programmed to vary phase shifts based on the wireless environment through electronic circuits integrated in the meta-surfaces.
We assume that, the channel matrix from the BS to the reflecting RIS, $\mathbf{H}_1 \in \mathbb{C}^{(N \times M)}$, and the channel vector $\mathbf{h}_{k,2} \in \mathbb{C}^{(N \times 1)}$ for all $k$, from the RIS to all the $K$ users are perfectly known at both the BS and the RIS, with the aid of the transmission of pilot signals and feedback channels. It should be noted that, obtaining CSI at the RIS is a challenging task, which definitely requires that the RIS has the capability to transmit and receive signals. However this is indeed contradictory to the claim that RIS does not need RF chains. One solution is to install RF chains dedicated to channel estimation. To this end, the system should be delicately designed to tradeoff the system performance and cost, which is beyond the scope of this paper.

Assume the frequency flat channel fading. The signal received at the $k^{th}$ user is given as
\begin{equation} \label{eq:sys_1}
\begin{split}
y_k=\mathbf{h}_{k,2}^T\mathbf{\Phi}\mathbf{H}_1\mathbf{Gx}+w_k, \\
\end{split}
\end{equation}
where $y_k$ denotes the signal received at the $k^{th}$ user. $\mathbf{x}$ is a column vector of dimension $K \times 1 $ consisting of data streams transmitted to all the users, with zero mean unit variance entries, $\mathcal{E}[|x|^2]=1$. $\mathbf{G} \in \mathbb{C}^{(M \times K)}$ is the beamforming matrix applied at the BS, while $\mathbf{\Phi}\triangleq\mathrm{diag}[\phi_1,\phi_2,\ldots,\phi_N], \in \mathbb{C}^{(N \times N)}$ is the phase shift matrix applied at the reflecting RIS. $w_k$ is the zero mean additive white Gaussian noise (AWGN) with entries of variance $\sigma_n^2$.

Note that, $\mathbf{\Phi}$ is a diagonal matrix whose entries are given by $\mathbf{\Phi}(n,n)=\phi_n=e^{j\varphi_n}$, where $\varphi_n$ is the phase shift induced by each element of the RIS. Here we assume ideal reflection by the RIS such that the signal power is lossless from each reflection element or $|\mathbf{\Phi}(n,n)|^2$=1. Then, the reflection results in the phase shift of the impinging signals only. In this paper, we consider the continuous phase shift where $\varphi_n \in [0,2\phi) \forall n$ for the development of DRL based algorithm. 

From (\ref{eq:sys_1}), it can be seen that, compared to MISO relaying systems, reflecting RIS assisted MISO systems do not introduce AWGN at the RIS. This is because that the RIS acts as a passive mirror simply reflecting the signals incident on it, without signal decoding and encoding. The phases of signals impinging on the RIS will be reconfigured through the micro-controller connected to the RIS. It is also clear that, the signals arriving at the users experience the composite channel fading, $\mathbf{h}_{k,2}^T\mathbf{\Phi}\mathbf{H}_1$. Compared to point to point wireless communications, this composite channel fading results in more severe signal loss, if without signal compensation at the RIS. 

To maintain the transmission power at the BS, the following constraint is considered
\begin{equation} \label{eq:sys_2}
\mathcal{E} \bigg \{ tr \{ \mathbf{Gx}(\mathbf{Gx})^{\mathcal{H}} \}  \bigg \} \leq P_t,
\end{equation}
where $P_t$ is the total transmission power allowed at the BS.

The received signal model (\ref{eq:sys_1}) can be further written as
\begin{equation} \label{eq:sys_1a}
\begin{split}
y_k=\mathbf{h}_{k,2}^T\mathbf{\Phi}\mathbf{H}_1\mathbf{g}_kx_k+\sum_{n,n\neq k}^K\mathbf{h}_{k,2}^T\mathbf{\Phi}\mathbf{H}_1\mathbf{g}_nx_n+w_k, \\
\end{split}
\end{equation}
where $\mathbf{g}_k$ is the $k^{th}$ column vector of the matrix $\mathbf{G}$.

Without joint detection of data streams for all users, the second term of (\ref{eq:sys_1a}) is treated as cochannel interference. The SINR at the $k^{th}$ user is given by
\begin{equation} \label{eq:sys_3}
\rho_{k}=\frac{|\mathbf{h}_{k,2}^T\mathbf{\Phi H}_1\mathbf{g}_k|^2}{\sum_{n, n\neq k}^K |\mathbf{h}_{k,2}^T\mathbf{\Phi H}_1\mathbf{g}_n|^2+\sigma_n^2}.
\end{equation}

In this paper, we adopt the ergodic sum rate, as given in (\ref{eq:sys_6}), as a metric to evaluate the system performance,
\begin{equation} \label{eq:sys_6}
C(\mathbf{G},\mathbf{\Phi}, \mathbf{h}_{k,2},\mathbf{H}_1)=\sum_{k=1}^K R_k,
\end{equation}
where $R_k$ is the data rate of the $k^{th}$ user, given by $R_k=\log_2(1+\rho_k)$. Unlike the traditional beamforming design and phase shift optimization algorithms that require full up-to-date cross-cell channel state information (CSI) for RIS-based systems. Our objective is to find out the optimal $\mathbf{G}$ and $\mathbf{\Phi}$ by maximizing sum rate $C$ leveraging the recent
advance of DRL technique under given a particular CSI. Unlike the conventional deep neural networks (DNN), where it needs two phases, offline training phase and online learning phase, our proposed DRL method, each CSI is used to construct the state, and run the algorithm to obtain the two matrices continuously. The optimization problem can be formulated as
\begin{equation} \label{eq:BD_1}
\begin{split}
& \max\limits_{\mathbf{G},\mathbf{\Phi}} \; C(\mathbf{G},\mathbf{\Phi}, \mathbf{h}_{k,2},\mathbf{H}_1)  \\
& \; \textrm{s.t.} \;\; tr\{\mathbf{G}\mathbf{G}^{\mathcal{H}} \} \leq P_t \\
& \;\;\;\quad\;\; |\phi_n|=1\;\forall n=1,2,\ldots,N.\\
\end{split}
\end{equation}
It can be seen that (\ref{eq:BD_1}) is a non-convex non-trivial optimization problem, due to the non-convex objective function and the constraint. Exhaustive search would have to be used to obtain the optimal solution if utilizing classical mathematical tools, which is impossible, particularly for large scale network. Instead, in general, algorithms are developed to find out suboptimal solutions employing alternating optimization techniques to maximize the objective functions, where in each iteration, suboptimal $\mathbf{G}$ is solved by first fixing $\mathbf{\Phi}$  \cite{Ref10a}-\cite{Ref10g} while suboptimal $\mathbf{\Phi}$ is derived by fixing the $\mathbf{G}$, until the algorithms converge. In this paper, rather than directly solving the challenging optimization problem mathematically, we formulate the sum rate optimization problem in the context of advanced DRL method to obtain the feasible $\mathbf{G}$ and $\mathbf{\Phi}$.

\section {Preliminary knowledge of DRL}
In this section, we briefly describe the background of DRL which builds up the foundation for the proposed joint design of transmit beamforming and phase shifts.
\begin{figure*}[ht]
\begin{center}
  \includegraphics[width=15.5cm]{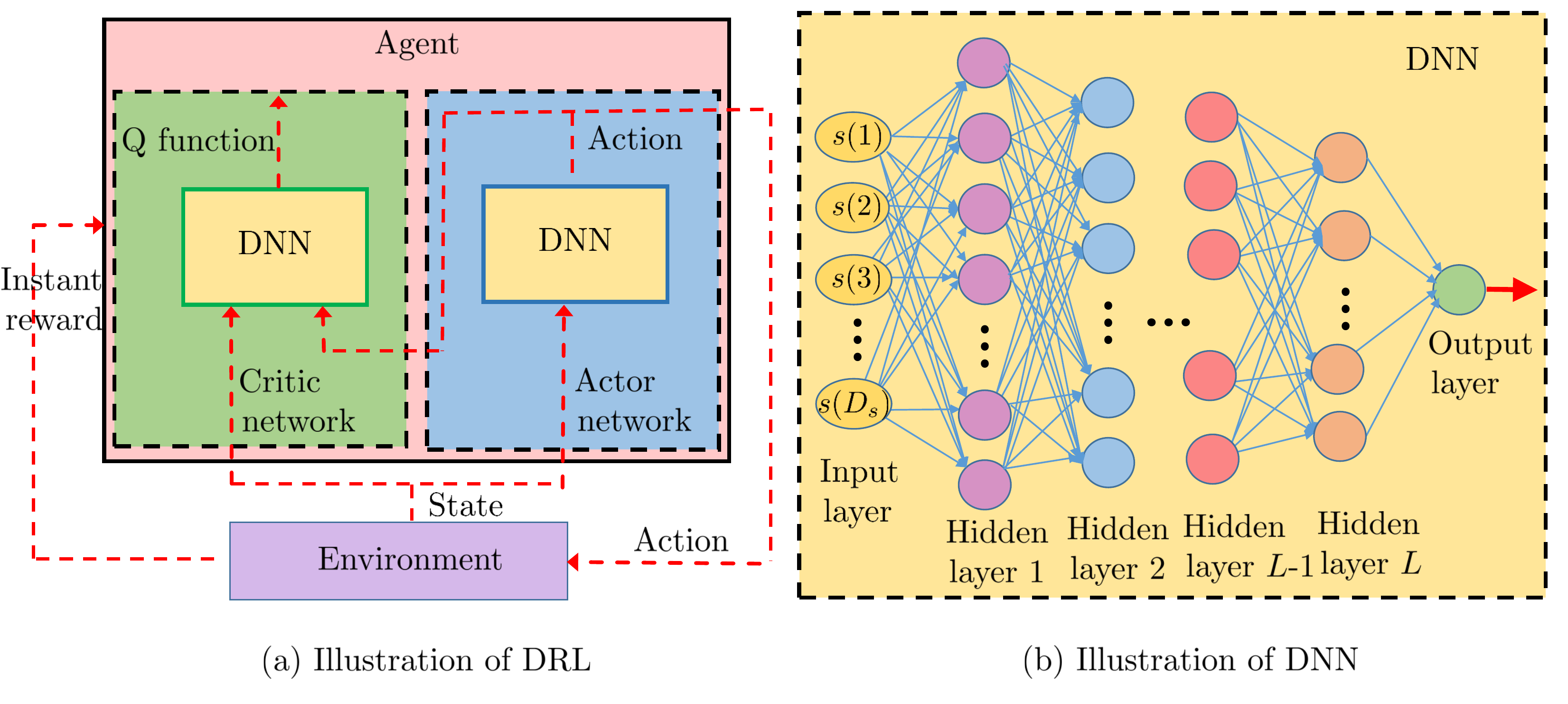}
  \caption{ (a) An illustration of deep Q learning, where a double DNN to approximate the optimal state-action value and Q function. (b) Illustration of the proposed DNN.}
  \label{fig:NN} \vspace{-8mm}
\end{center}
\end{figure*}

\subsection {Overview of DRL}
In a typical RL, the agent gradually derives its best action through the trial-and-error interactions with the environment over time, applying actions to the environment, observing the instant rewards and the transitions of state of the environment, as shown in Fig. \ref{fig:NN}. There are a few basic elements used to fully characterize the RL learning process, the state, the action, the instant reward, the policy and the value function:

\emph{(1) State:} a set of observations characterizing the environment. The state $s^{(t)} \in \textit{S}$ denotes the observation at the time step $t$.

\emph{(2) Action:} a set of choices. The agent takes one action step by step during the learning process. Once the agent takes an action $a^{(t)} \in \textit{A}$ at time instant $t$ following a policy $\pi$, the state of the environment will transit from the current state $s^{(t)}$ to the next state $s^{(t+1)}$. As a result, the agent gets a reward $r^{(t)}$.

\emph{(3) Reward:} a return. The agent wants to acquire a reward by taking action $a$ given state $s$. It is also a performance metric $r^{(t)}$ to evaluate how good the action $a^{(t)}$ is given a state $s^{(t)}$ at time instant $t$.

\emph{(4) Policy:} the policy $\pi(s^{(t)},a^{(t)})$ denotes the probability of taking action $a^{(t)}$ conditioned on the state $s^{(t)}$. Note that, the policy function satisfies $\sum_{a^{(t)} \in \textit{A}}\pi(s^{(t)},a^{(t)})=1$.

\emph{(5) State-action value function:} value to be in a state $s$ and action $a$. The reward measures immediate return from action $a$ given state $s$, whereas the value function measures potential future rewards which the agent may get from taking action $a$ being in the state $s$.

\emph{(6) Experience:} defined as $(s^{(t)},a^{(t)}, r^{(t+1)}, s^{(t+1)})$.

Adopting the $Q_{\pi}(s^{(t)},a^{(t)})$ function as the state-action value function. Given the state $s^{(t)}$, the action $a^{(t)}$, and the instant reward $r^{(t)}$ at time $t$, the $Q$ value function is given as
\begin{equation} \label{eq:BD_31}
\begin{split}
&Q_{\pi}(s^{(t)},a^{(t)})=\mathcal{E}_{\pi}\big [ R^{(t)}|s^{(t)}=s, a^{(t)}=a \big ] \\
&R^{(t)}=\sum_{\tau=0}^{\infty} \gamma^{\tau} r^{(t+\tau +1)},
\end{split}
\end{equation}
where $\gamma \in (0,1]$ is the discount rate. The $Q$ function is a metric to evaluate the impact of the choice of action on the expected future cumulative discounted reward achieved by the learning process, with the choice of action $a^{(t)}$ under the policy $\pi$.

The $Q$ function satisfies the Bellman equation given by
\begin{equation} \label{eq:BD_31a}
\begin{split}
Q_{\pi}(s^{(t)},a^{(t)})=&\mathcal{E}_{\pi}\big [ r^{(t+1)} |s^{(t)}=s, a^{(t)}=a \big ]+\\
&\gamma \sum_{s' \in \textit{S}}P_{ss'}^a \bigg (\sum_{a' \in \textit{A}} \pi(s', a')Q^{\pi}(s',a') \bigg),
\end{split}
\end{equation}
where $P_{ss'}^a=P_r(s^{(t+1)}=s'|s^{(t)}=s,a^{(t)}=a)$ is the transition probability from state $s$ to state $s'$ with action $a$ being taken.

The Q-learning algorithm searches the optimal policy $\pi^*$. From (\ref{eq:BD_31a}), the optimal $Q$ function associated with the optimal policy becomes
\begin{equation} \label{eq:BD_31b}
\begin{split}
Q^*(s^{(t)},a^{(t)})=&r^{(t+1)}(s^{(t)}=s, a^{(t)},\pi=\pi^*) +\\
&\gamma \sum_{s' \in \textit{S}}P_{ss'}^a \max_{a' \in \textit{A}} Q^*(s',a').
\end{split}
\end{equation}

The Bellman equation (\ref{eq:BD_31b}) can be solved recursively to obtain the optimal $Q^*(s^{(t)},a^{(t)})$, without the knowledge of exact reward model and the state transition model. The updating on the Q function is given as
%
\begin{equation} \label{eq:BD_32a}
\begin{split}
Q^{*}(s^{(t)},a^{(t)})\leftarrow &(1-\alpha)Q^{*} (s^{(t)},a^{(t)})+ \alpha (r^{(t+1)}+\\
&\gamma \max_{a'}Q_{\pi}(s^{(t+1)},a')),
\end{split}
\end{equation}
where $\alpha$ is the learning rate for the update of $Q$ function.

If $Q(s(t),a(t))$ is updated at every time instant, it will converge to the optimal state-action value function $Q^*(s(t),a(t))$. However, this is not easily achieved, particularly with the large dimension state space and action space. Instead, the function approximation is usually used to address problems with enormous state/action spaces. Popular function approximators include feature representation, neural networks and ones directly relating value functions to state variables. Rather than utilizing explicit mathematical modeling, the DNN approximates state/action value function, policy function and the system model as composition of many non-linear functions, as shown in Fig. \ref{fig:NN}, where both the Q function and the action are approximated by DNN. However, the neural network based approximation does not give any interpretation and the resulting DRL based algorithm might also give local optimal due to the sample correlation and non-stationary targets.

One key issue using neural network as Q function approximation is that, the states are highly correlated in time domain and result in reduction of randomness of states since they are all extracted from the same episode. The experience replay, which is a buffer window consisting of last few states, can considerably improve the performance of DRL. Instead of updating from the last state, the DNN updates from a batch of randomly sampled states to the experience replay.

With DRL, the $Q$ value function is completely determined by a parameter vector $\mathbf{\theta}$
\begin{equation} \label{eq:BD_32b}
\begin{split}
Q(s(t),a(t))\triangleq Q(\mathbf{\theta} |s(t),a(t)),
\end{split}
\end{equation}
where $\mathbf{\theta}$ is equivalent to the weighting and bias parameters in the neural network. Rather than updating the $Q$ function directly as in (\ref{eq:BD_31a}), with DRL, the optimal Q value function can be approached by updating $\mathbf{\theta}$ using stochastic optimization algorithms
\begin{equation} \label{eq:BD_32c}
\mathbf{\theta}^{(t+1)}=\mathbf{\theta}^{(t)}-\mu \Delta_{\mathbf{\theta}} \ell (\mathbf{\theta}),
\end{equation}
where $\mu$ is the learning rate for the update on $\mathbf{\theta}$ and $\Delta_{\mathbf{\theta}}$ is the gradient of loss function $\ell (\mathbf{\theta})$ with respect to $\mathbf{\theta}$.

The loss function is generally given as the difference between the neural network predicted value and the actual target value. However, since reinforcement learning is a process learning to approach the optimal $Q$ value function, the actual target value is not known. To address this problem, two neural networks with the identical architecture are defined, the training neural network and the target neural network, whose value functions are respectively given by $Q(\mathbf{\theta}^{(train)}|s^{(t)},a^{(t)})$ and $Q(\mathbf{\theta}^{(target)}|s^{(t)},a^{(t)})$. The target neural network is synchronized to the training neural network at a predetermined frequency. The actual target value is estimated as
\begin{equation} \label{eq:BD_32c1}
\begin{split}
y=r^{(t+1)}+\gamma \max_{a'} Q(\mathbf{\theta}^{(target)}|s^{(t+1)},a').
\end{split}
\end{equation}
The loss function is thus given by
\begin{equation} \label{eq:BD_32c2}
\begin{split}
&\ell (\mathbf{\theta})=\bigg ( y- Q(\mathbf{\theta}^{(train)}|s^{(t)},a^{(t)}) \bigg )^2.
\end{split}
\end{equation}

\subsection{DDPG}
As the proposed joint design of transmit beamforming and phase shifts is casted as a DRL optimization problem, the most challenging of it is the continuous state space and action space. To address this issue, we explore the DDPG neural network to solve our optimization problem, as shown in Fig. \ref{fig:NN}. It can be seen that, there are two DNNs in DDPG neural network, the actor network and the critic network. The actor network takes the state as input and outputs the continuous action, which is in turn input to the critic network together with the state. The actor network is used to approximate the action, thus eliminating the need of finding the action maximizing the Q value function given the next state which involves non-convex optimization.

The updates on the training critic network are given as follows:
\begin{equation} \label{eq:BD_32d}
\begin{split}
&\mathbf{\theta}_c^{(t+1)}=\mathbf{\theta}_c^{(t)}-\mu_c \Delta_{\mathbf{\theta}_c^{(train)}} \ell (\mathbf{\theta}_c^{(train)}),
\end{split}
\end{equation}

\begin{equation} \label{eq:BD_32d1}
\begin{split}
\ell (\mathbf{\theta}_c^{(train)})=&\bigg ( r^{(t)}+\gamma q(\mathbf{\theta}_c^{(target)}|s^{(t+1)},a')-\\ &q(\mathbf{\theta}_c^{(train)}|s^{(t)},a^{(t)}) \bigg )^2,
\end{split}
\end{equation}
where $\mu_c$ is the learning rate for the update on training critic network. $a'$ is the action output from the target actor network and $\Delta_{\mathbf{\theta}_c^{(train)}} \ell (\mathbf{\theta}_c^{(train)})$ denotes the gradient with respect to the training critic network $\mathbf{\theta}_c^{(train)}$. The $\mathbf{\theta}_c^{(target)}$ and the $\mathbf{\theta}_c^{(train)}$ denote the training and the target critic network, in which the parameters of the target network are updated as that of the training network in certain time slots. The update on target network is much slower than the training network. The update on the training actor network is given as
\begin{equation} \label{eq:BD_32e}
\begin{split}
&\mathbf{\theta}_a^{(t+1)}\!=\!\mathbf{\theta}_a^{(t)}\!-\!\mu_a \Delta_{a}q(\mathbf{\theta}_c^{(target)}|s^{(t)},a) \Delta_{\mathbf{\theta}_a^{(train)}}\pi (\mathbf{\theta}_a^{(train)}|s^{(t)})
\end{split}
\end{equation}
where $\mu_a$ is the learning rate for the update on training actor network. $\pi (\mathbf{\theta}_a^{(train)}|s^{(t)})$ denotes the training actor network with $\mathbf{\theta}_a^{(train)}$ being the DNN parameters and given input $s^{(t)}$. $\Delta_{a}q(\mathbf{\theta}_c^{(target)}|s^{(t)},a)$ is the gradient of target critic network with respect to the action, whereas $\Delta_{\mathbf{\theta}_a^{(train)}}\pi (\mathbf{\theta}_a^{(train)}|s^{(t)})$ is the gradient of training actor network with respect to its parameter $\mathbf{\theta}_a^{(train)}$. It can be seen from (\ref{eq:BD_32e}), the update of training actor network is affected by the target critic network through gradient of the target critic network with respect to the action, which ensures that the next selection of action is on the favorite direction of actions to optimize the $Q$ value function.

The updates on the target critic network and the target actor network are given as follows, respectively
\begin{equation} \label{eq:BD_32f}
\begin{split}
& \mathbf{\theta}_c^{(target)}\leftarrow \tau_c \mathbf{\theta}_c^{(train)}+(1-\tau_c)\mathbf{\theta}_c^{(target)}, \\
& \mathbf{\theta}_a^{(target)}\leftarrow \tau_a \mathbf{\theta}_a^{(train)}+(1-\tau_a)\mathbf{\theta}_a^{(target)},\\
\end{split}
\end{equation}
where $\tau_c, \tau_a$ are the learning rate for updating of the target critic network and the target actor network, respectively.

\section {DRL based Joint Design of Transmit Beamforming and Phase Shifts}
\begin{figure}
\begin{center}
  \includegraphics[width=9.5cm]{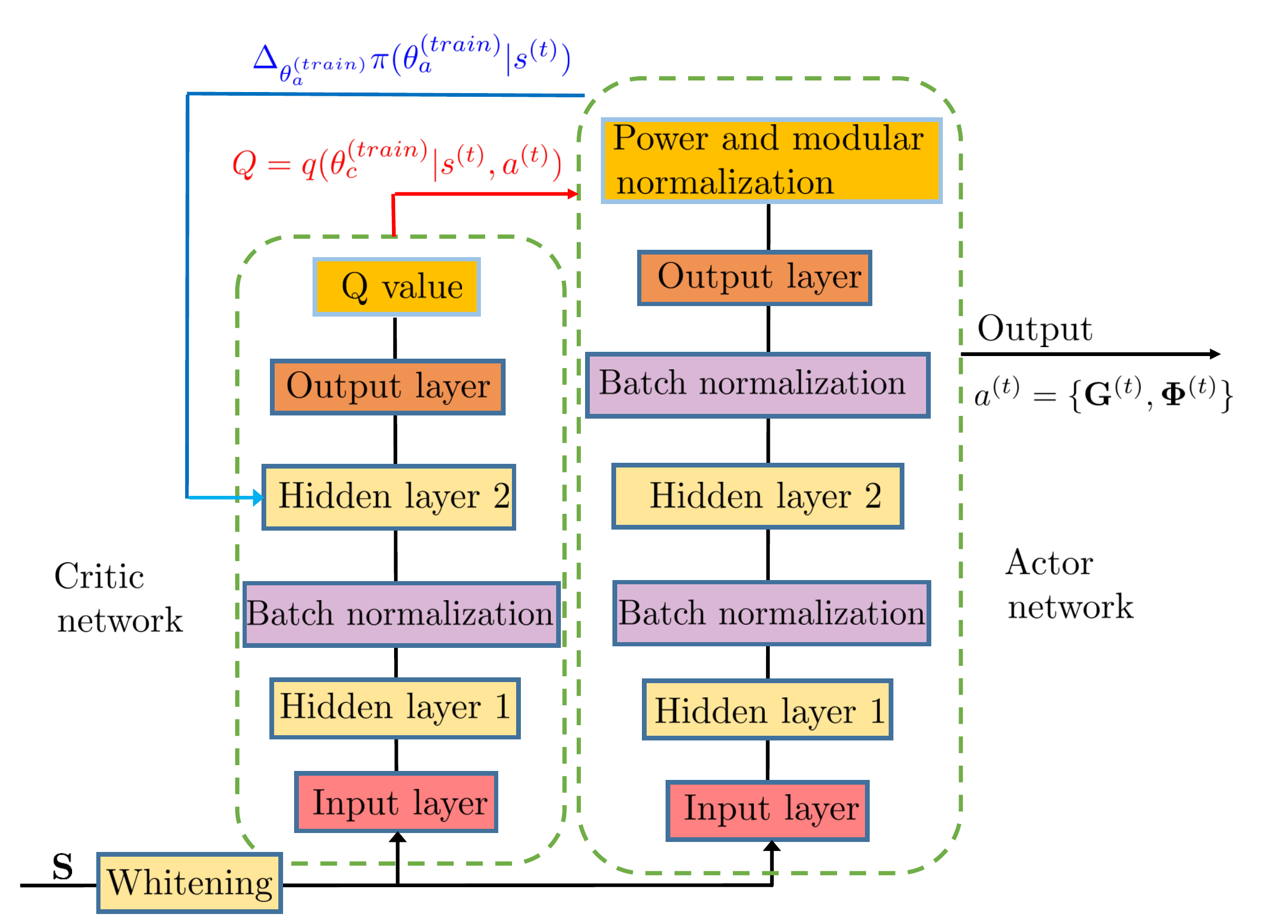}
  \caption{Proposed DNN structure of the critic network
and the actor network.}
  \label{fig:layout}
\end{center}  \vspace{-8mm}
\end{figure}

In this section, we present the proposed DRL based algorithm for joint design of transmit beamforming and phase shifts, utilizing DDPG neural network structure shown in Fig. \ref{fig:layout}. The DRL algorithm is driven by two DNNs, the state $s$, the action $a$ and the instant reward $r$. First we introduce the structure of the proposed DNNs, followed by detailed description of $s, a, r$ and the algorithm.

\subsection{Construction of DNN}
The structures of DNN utilized in this paper are shown in Fig. \ref{fig:layout}. As can be seen, both proposed DNN structures of the critic network and the actor network of are a fully connected deep neural network. The critic network and the actor network have the identical structure, comprised of one input layer, one output layer, and two hidden layers. The input and the output dimension of the critic network equals to the cardinality of the state set together with the action set and the $Q$ value function, respectively. The input and output dimension of the actor network are defined as the cardinality of the state and the action, respectively. The number of neurons of hidden layers depend on the number of users, the number of antennas at the BS and the number of elements at RIS. In general, the number of neurons of hidden layers must be larger than the input and the output dimension. The action output from the actor network will be input to the hidden layer 2 to avoid Python implementation issues in the computation of $\Delta_{a}q(\mathbf{\theta}_c^{(target)}|s^{(t)},a)$.

Note that, the correlation between entries of $s$ will degrade the efficiency of using neural network as function approximation. To overcome this problem, prior to being input to both the critic and the actor network, the state $s$ will go through a whitening process, to remove the correlation between the entries of the state $s$.

In order to overcome the variation on distribution of each layer's inputs resulting from the changes in parameters of the previous layers, batch normalizing is utilized at the hidden layers. Batch normalization allows for much higher learning rates and less careful about initialization, and in some cases eliminates the need for dropout.

The activation function utilized here is $tanh$ in order to address the negative inputs. The optimizer used for both the training critic network and the training actor network is Adam with adaptive learning rate $\mu_c^{(t)}=\lambda_c\mu_c^{(t-1)}$ and $\mu_a^{(t)}=\lambda_a\mu_a^{(t-1)}$, where $\lambda_c$ and $\lambda_a$ are the decaying rate for the training critic network and training actor network.

Noting that, $\mathbf{G}$ should satisfied the power constraint defined in (\ref{eq:BD_1}). To implement this, a normalization layer is employed at the output of the actor network, where $Tr \{\mathbf{G}\mathbf{G}^T \} =P_t$. For $\mathbf{\Phi}$, $|\mathbf{\Phi}(n,n)|^2=1$  be maintained to ensure signal reflection without the power consumption.

\subsection{Algorithm Description}
Assuming there exists a central controller, or the agent, which is able to instantaneously collect the channel information, $\mathbf{H}_1$ and $\mathbf{h}_{k,2} \forall k$. At time step $t$, given the channel information, and the action $\mathbf{G}^{(t-1)}$ and $\mathbf{\Phi}^{(t-1)}$ in the previous state, the agent constructs the state $s^{(t)}$ for time step $t$ following the section IV.B.1 State.

At the beginning of the algorithm, the experience replay buffer $\mathcal{M}$, the critic network and the actor network paramters $\mathbf{\theta}_c^{(train)}$ and $\mathbf{\theta}_a^{(train)}$, the action $\mathbf{G}$ and $\mathbf{\Phi}$ need to be initialized. In this paper, we simply adopt the identity matrix to initialize $\mathbf{G}$ and $\mathbf{\Phi}$.

\begin{algorithm}
\caption{Joint Transmit beamforming and Phase Shifts Design}
\label{alg:ALG1}
\textbf{Input:} $\mathbf{H}_1$, $\mathbf{h}_{k,2}, \forall k$ \\
\textbf{Output:} optimal action $a=\{ \mathbf{G}, \mathbf{\Phi} \}$, $Q$ value function\\
\textbf{Initialization:} experience replay memory $\mathcal{M}$ with size $D$, training actor network parameter $\mathbf{\theta}_a^{(train)}$, target actor network parameter $\mathbf{\theta}_a^{(train)}=\mathbf{\theta}_a^{(target)}$, training critic network with parameter $\mathbf{\theta}_c^{(train)}$, target critic network with parameter $\mathbf{\theta}_c^{(train)}=\mathbf{\theta}_c^{(target)}$, transmit beamforming matrix $\mathbf{G}$, phase shift matrix $\mathbf{\Phi}$ \\
\textbf{Do:}
\begin{algorithmic}[1]
\FOR{\texttt{espisode $=0,1,2, \cdots,N-1$}}
\STATE Collect and preprocess $\mathbf{H}_1^{(n)}, \mathbf{h}_{k,2}^{(n)}, \forall k$ for the $n^{th}$ episode to obtain the first state $s^{(0)}$
\FOR{\texttt{t=$0,1,2, \cdots, T-1$}}
\STATE Obtain action $a^{(t)}=\{\mathbf{G}^{(t)},\mathbf{\Phi}^{(t)} \}=\pi(\mathbf{\theta}_a^{(train)})$ from the actor network \\
\STATE\hspace{\algorithmicindent} Observe new state $s^{(t+1)}$ given action $a^{(t)}$ \\
\STATE\hspace{\algorithmicindent} Observe instant reward $r^{(t+1)}$ 
\STATE\hspace{\algorithmicindent} Store the experience $(s^{(t)}, a^{(t)}, r^{(t+1)}, s^{(t+1)})$ in \STATE\hspace{\algorithmicindent} the replay memory \\
\STATE Obtain the $Q$ value function as $Q=q(\mathbf{\theta}_c^{(train)}|s^{(t)}, a^{(t)})$ from the critic network\\
\STATE Sample random mini-batches of size $W$ of experiences from replay memory $\mathcal{M}$\\
\STATE Construct training critic network loss function $\ell(\mathbf{\theta}_c^{(train)})$ given by (\ref{eq:BD_32d1})\\
\STATE Perform SGD on training critic network to obtain $\Delta_{\mathbf{\theta}_c^{(train)}}\ell (\mathbf{\theta}_c^{(train)})$\\
\STATE Perform SGD on target critic network to obtain $\Delta_{a}q(\mathbf{\theta}_c^{(target)}|s^{(t)},a)$\\
\STATE Perform SGD on training actor network to obtain $\Delta_{\mathbf{\theta}_a^{(train)}}\pi(\mathbf{\theta}_a^{(train)}|s^{(t)})$ \\
\STATE Update the training critic network $\mathbf{\theta}_c^{(train)}$ \\
\STATE Update the training actor network $\mathbf{\theta}_a^{(train)}$ \\
\STATE Every $U$ steps, update the target critic network $\mathbf{\theta}_c^{(target)}$ \\
\STATE Every $U$ steps, update the target actor network $\mathbf{\theta}_a^{(target)}$ \\
\STATE Set input to DNN as $s^{(t+1)}$
\ENDFOR
\ENDFOR \vspace{-0mm}
\end{algorithmic}  \vspace{-0mm}
\end{algorithm} \vspace{-0mm}

The algorithm is run over $N$ episodes and each episode iterates $T$ steps. For each episode, the algorithm terminates whenever it converges or reaches the maximum number of allowable steps. The optimal $\mathbf{G}_{opt}$ and $\mathbf{\Phi}_{opt}$ are obtained as the action with the best instant reward. Note that the purpose of this algorithm is to obtain the optimal $\mathbf{G}$ and $\mathbf{\Phi}$ utilizing DRL, rather than to train a neural network for online processing. The details of the proposed method are shown in Algorithm \ref{alg:ALG1}.

The construction of the state $s$, the action $\mathbf{G}, \mathbf{\Phi}$,  and the instant reward are described in details as follows.

\subsubsection{State}
The state $s^{(t)}$ at the time step $t$ is determined by the transmission power at the $t^{th}$ time step, the received power of users at the $t^{th}$ time step, the action from the $(t-1)^{th}$ time step, the channel matrix $\mathbf{H}_1$ and $\mathbf{h}_{k,2} \in k$. Since the neural network can only take real rather than complex numbers as input, in the construction of the state $s$, if a complex number is involved, the real part and the imaginary part will be separated as independent input port. Given transmit symbols with unit variance, the transmission power for the $k^{th}$ user is given by $||\mathbf{G}_k||^2=|Re\{\mathbf{G}_k^{\mathcal{H}}\mathbf{G}_k\}|^2+|Im\{\mathbf{G}_k^{\mathcal{H}}\mathbf{G}_k\}|^2$. The first term is the contribution from the real part, whereas the second term is the contribution from the imaginary part, both of which are used as the independent input port to the critic network and the actor network. In total, there will be $2K$ entries of the state $s$ formed by the transmission power. Assuming that $\mathbf{\tilde{h}}_{k,2}=\mathbf{h}_{k,2}^T\mathbf{\Phi}\mathbf{H}_1\mathbf{G}$. The received power at the $k^{th}$ user contributed by the $k_2^{th}$ user is given as $|\mathbf{\tilde{h}}_{k,2}(n)|^2=|Re\{ \mathbf{\tilde{h}}_{k,2}(n) \}|^2+|Im\{ \mathbf{\tilde{h}}_{k,2}(n)\}|^2$. Likewise, both the power contributed by the real part and the imaginary part are used as independent input port to the critic network and the actor network. The total number of entries formed here is $2K^2$. The real part and the imaginary part of each entry of $\mathbf{H}_1$ and $\mathbf{h}_{k,2} \in k$ are also used as entries of the state. The total number of entries of the state constructed from the action at the $(t-1)^{th}$ step is given by $2MK+2N$, while the total number of entries from $\mathbf{H}_1$ and $\mathbf{h}_{k,2} \forall k$ is $2NM+2KN$.

In summary, the dimension of the state space is $D_s=2K+2K^2+2N+2MK+2NM+2KN$. The reason we differentiate the transmission power and the receiving power contributed by the real part and the imaginary part is that, both the $\mathbf{G}$ and $\mathbf{\Phi}$ are matrix with complex entries and the transmission and receiving power only will result in information lost due to the absolute operator.


\subsubsection{Action}
The action is simply constructed by the transmit beamforming matrix $\mathbf{G}$ and the phase shift matrix  $\mathbf{\Phi}$. Likewise, to tackle with the real input problem, $\mathbf{G}=Re\{\mathbf{G} \}+Im \{ \mathbf{G}\}$ and $\mathbf{\Phi}=Re \{\mathbf{\Phi} \} +Im \{\mathbf{\Phi} \}$ are separated as real part and imaginary part, both are entries of the action. The dimension of the action space is $D_a=2MK+2N$.

\subsubsection{Reward}
At the $t^{th}$ step of the DRL, the reward is determined as the sum rate capacity $C(\mathbf{G}^{(t)},\mathbf{\Phi}^{(t)}, \mathbf{h}_{k,2},\mathbf{H}_1)$, given the instantaneous channels $\mathbf{H}_1$,  $\mathbf{h}_{k,2}, \forall k$ and the action $\mathbf{G}^{(t)}$ and $\mathbf{\Phi}^{(t)}$ obtained from the actor network.
\begin{table}
\caption{Hyper-parameters Descriptions} \label{tab:hyperP} \vspace{-4mm}
\begin{center} \vspace{-4mm}
\begin{tabular}{ | m{3.5em} | m{15em}| m{3em} | }
\hline
Parameter & description & value \\
\hline
$\gamma$ & discounted rate for future reward & 0.99 \\
\hline
$\mu_c$ & learing rate for training critic network uptate & 0.001 \\
\hline
$\mu_a$ & learing rate for training actor network uptate & 0.001 \\
\hline
$\tau_c$ & learing rate for target critic network uptate & 0.001 \\
\hline
$\tau_a$ & learing rate for target actor network uptate & 0.001 \\
\hline
$\lambda_c$ & decaying rate for training critic network uptate & 0.00001 \\
\hline
$\lambda_a$ & decaying rate for training actor network uptate & 0.00001 \\
\hline
$D$ & buffer size for experience replay& 100000 \\
\hline
$N$ & the number of episodes & 5000 \\
\hline
$T$ & the number of steps in each episode & 20000 \\
\hline
$W$ & the number of experiences in the mini-batch & 16 \\
\hline
$U$ & the number of steps synchronizing target network with the training network & 1 \\
\hline
\end{tabular} \vspace{-4mm}
\end{center}
\end{table}

\section{Numerical Results and Analysis}

In this section, we present performance evaluation for the proposed DRL based algorithm. In the simulations, we randomly generate channel matrix $\mathbf{H}_1$ and $\mathbf{h}_{k,2}, \forall k$ following rayleigh distribution. We assume that, the large scale path loss and the shadowing effects have been compensated. This is because the objective of this paper in its current format is to develop a framework for the optimal beamforming design and phase shift matrices by employing advanced DRL technique. Once the framework is ready, the effects of the path loss, the shadowing effects, the distribution of users and the direct link from the BS to the users can be easily investigated, through scaling DNNs, reconstructing the state, the action and the reward. All presented illustrations have been averaged results over 500 independent realizations.
\subsection{Setting and benchmarks}
The hyper-parameters used in the algorithm are shown in Table \ref{tab:hyperP}. We select two state-of-the-art algorithms as benchmarks. These are the weighted minimum mean square error (WMMSE) algorithm \cite{WMMSE}-\cite{WMMSEa} and an iterative algorithm based on fractional programming (FP) with the ZF beamforming \cite{chongwentwc2019}. In their generic forms, both algorithms require full up-to-date cross-cell CSI. Both are centralized and iterative in their original forms. The iterative FP algorithm with the ZF beamforming used in this paper is formulated in \cite{chongwentwc2019} Algorithm 3. Similarly, a detailed explanation and pseudo code of the WMMSE algorithm is given in \cite{WMMSEa} Algorithm 1. The performance of the proposed DRL-based algorithm in comparison with these state-of-the-art benchmarks is illustrated in the following.
\begin{figure}[htbp]
\begin{center}
  \includegraphics[width=9.6cm]{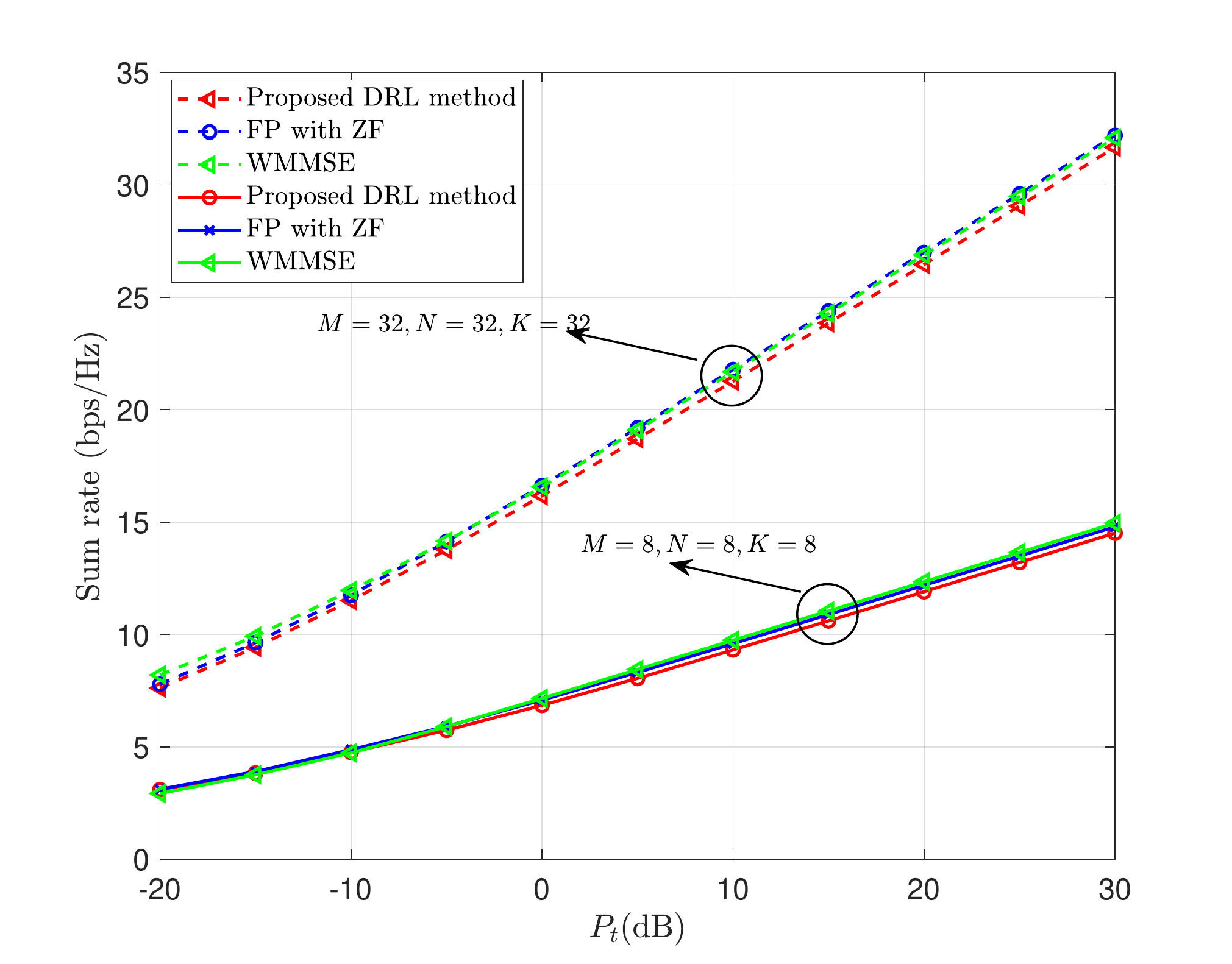} \vspace{-2mm}
  \caption{Sum rate versus $P_t$ to show the proposed DRL-based algorithm in comparison with two benchmarks.}
  \label{fig:comparison}
\end{center} \vspace{-6mm}
\end{figure}
\begin{figure}[ht]
\begin{center}
  \includegraphics[width=9.5cm]{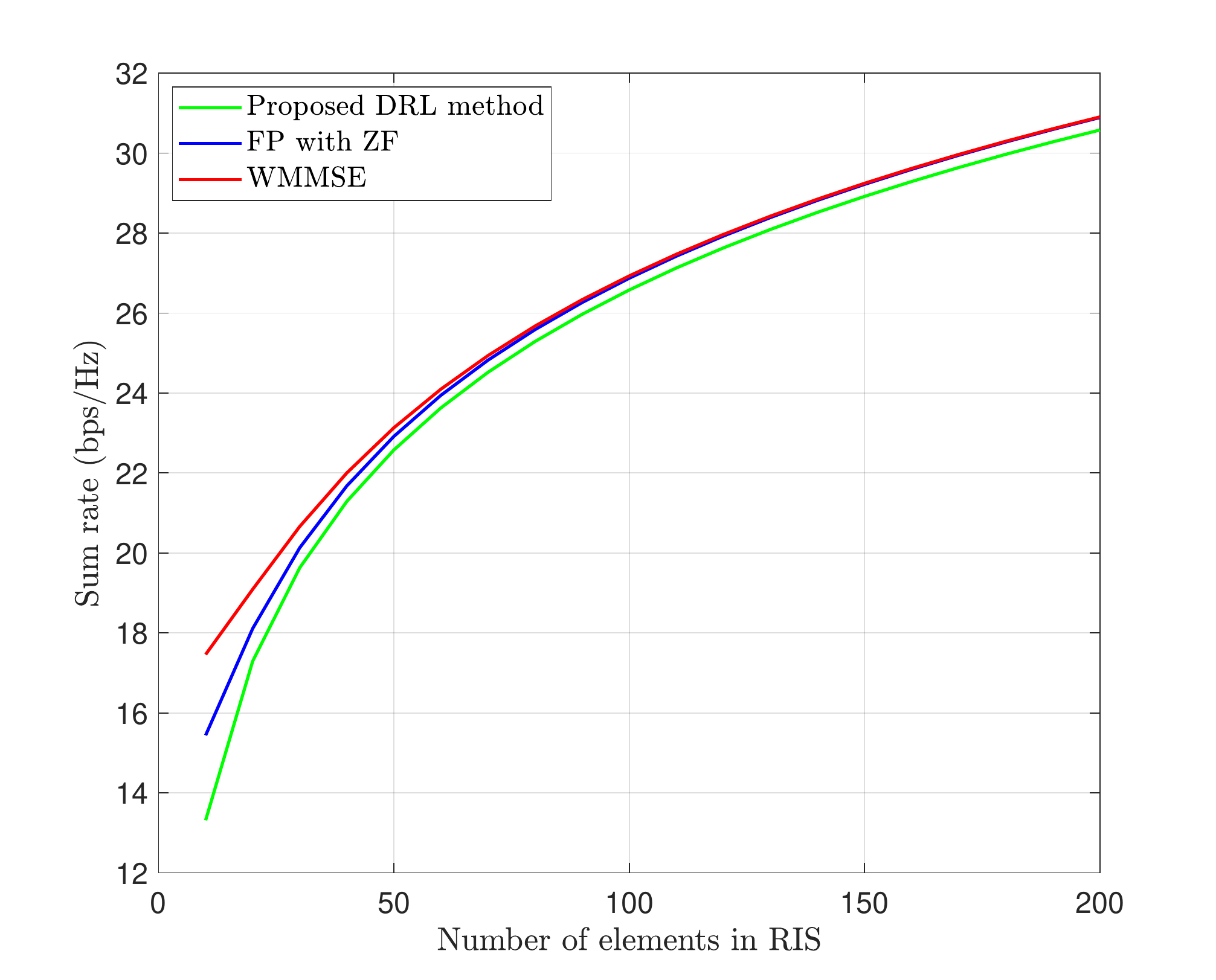}
  \caption{Sum rate as a function of element number $N$ with the proposed DRL-based algorithm as well as two benchmarks, $P_t=20dB, M=64, K=64$.}
  \label{fig:sumrateN}
\end{center}  \vspace{-6mm}
\end{figure}
\subsection{Comparisons with Benchmarks}
We have evaluated the proposed DRL-based approach described in Algorithm 1 as well as two benchmarks. Fig. \ref{fig:comparison} shows the sum rate versus maximum transmit power $P_t$. We consider two sets of system parameters, namely $M=32, N=32, K=32$, and  $M=8, N=8, K=8$.  It can be seen that our proposed DRL-based algorithm obtains the comparable sum-rate performance with these state-of-the-art benchmarks (WMMSE and FP optimization algorithm with ZF), and the sum rates increase with the transmit power $P_t$ under all considered algorithms and scenarios.

To further verify our proposed algorithm in more wider application scenarios, we perform another a simulation, which compares the sum rate as a function of the number of elements in RIS $N$ shown in Fig. \ref{fig:sumrateN} for $P_t=20dB, M=64, K=64$. It is observed that, the average sum rates increase with the $N$, resulting from the increase in the sum power of reflecting RIS as $N$ increases. This is achieved at the cost of the complexity of implementing RIS. It also further indicates that our proposed algorithm is robust in considered wider application scenarios, and approaching the optimal performance.
\subsection{Impact of $P_t$ on DRL}

\begin{figure}
\begin{center}
  \includegraphics[width=9.1cm]{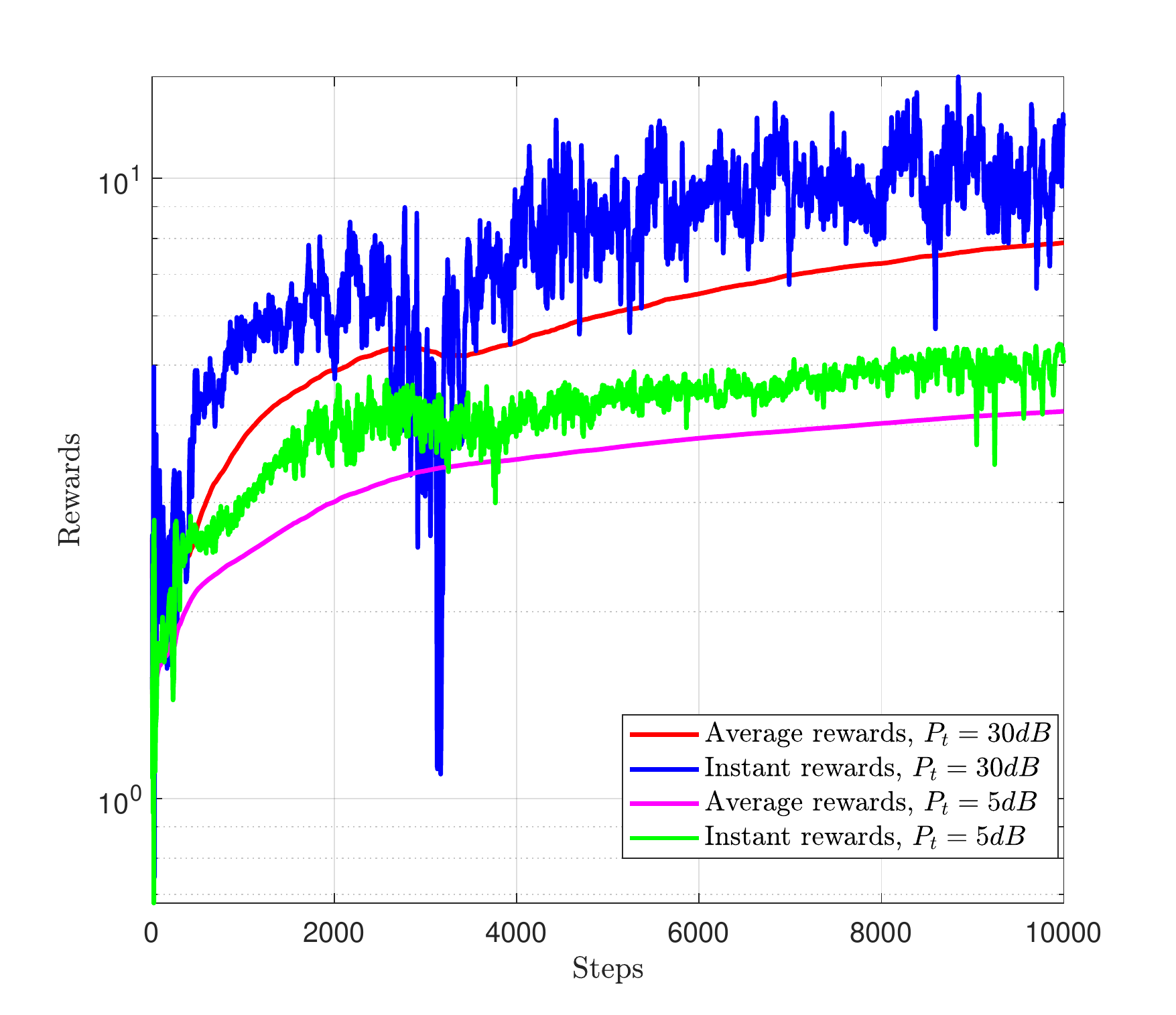} \vspace{-2mm}
  \caption{Rewards as a function of time steps at $P_t=5dB$ and $P_t=20dB$ respectively.}
  \label{fig:steps5dB} \vspace{-6mm}
\end{center}
\end{figure}
\begin{figure}[htbp]
\begin{center}
  \includegraphics[width=9.5cm]{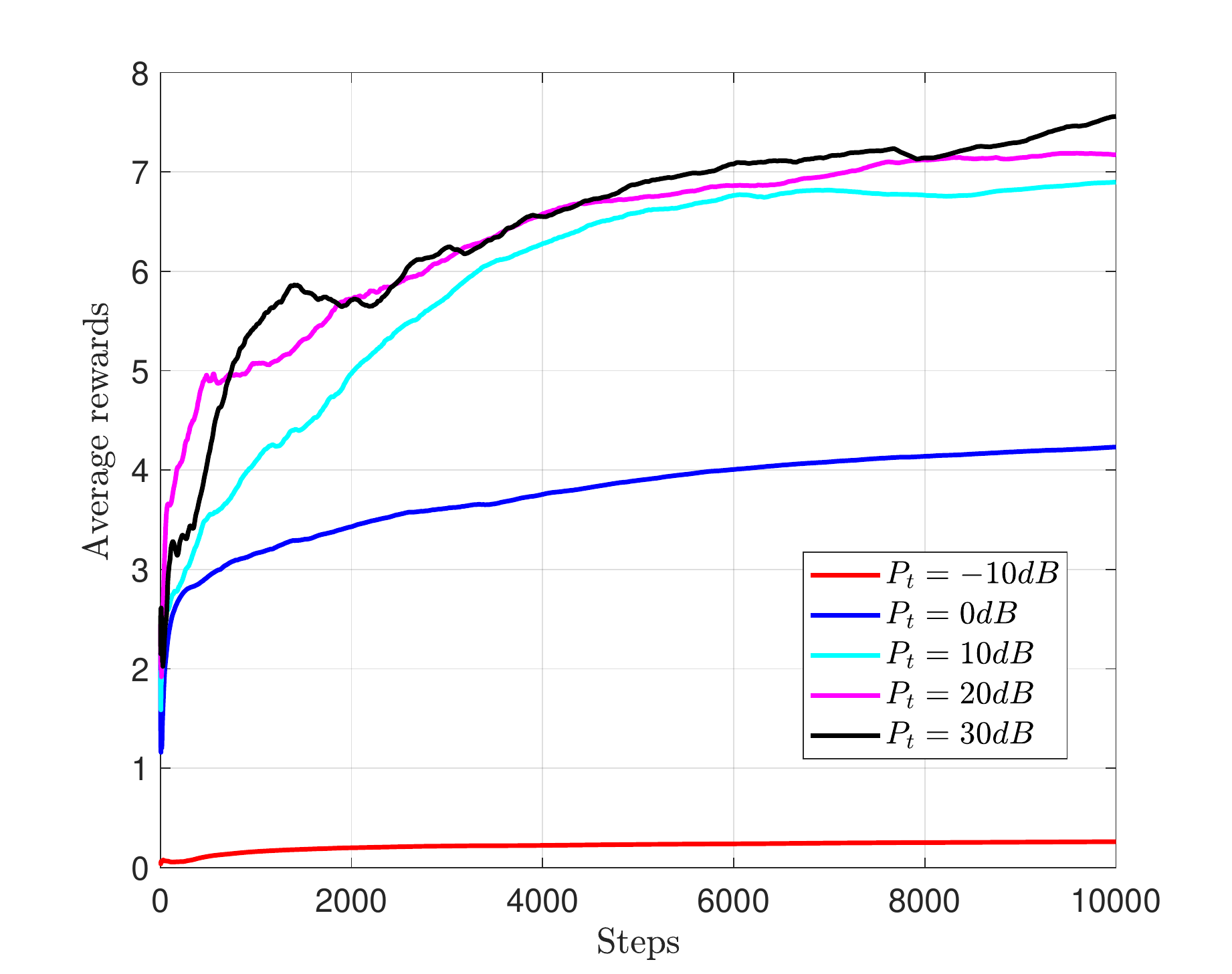}
  \caption{Average rewards versus time steps under different $P_r=\{-10dB, 0dB, 10dB, 20dB, 30dB\}$.}
  \label{fig:rewardpt}
\end{center} \vspace{-6mm}
\end{figure}

To get better understanding of our proposed DRL-based method, we investigate the impact of $P_t$ on it shown in Fig. \ref{fig:steps5dB}, in which we considered two settings $P_t=5dB$ and $P_t=20dB$ with rewards (instant rewards and average rewards)  as a function of time steps. In simulations, we use the following method to calculate the average rewards,
\begin{align} \label{Prob:Stepsize1}
\textrm{average}\_\textrm{reward}(K_i)=\frac{\sum_{k=1}^{K_i} \textrm{reward} (k)}{K_i}, K_i=1,2,...,K,
\end{align}
where $K$ is the maximum steps.
\begin{figure}[htbp]
\begin{center}
  \includegraphics[width=9.6cm]{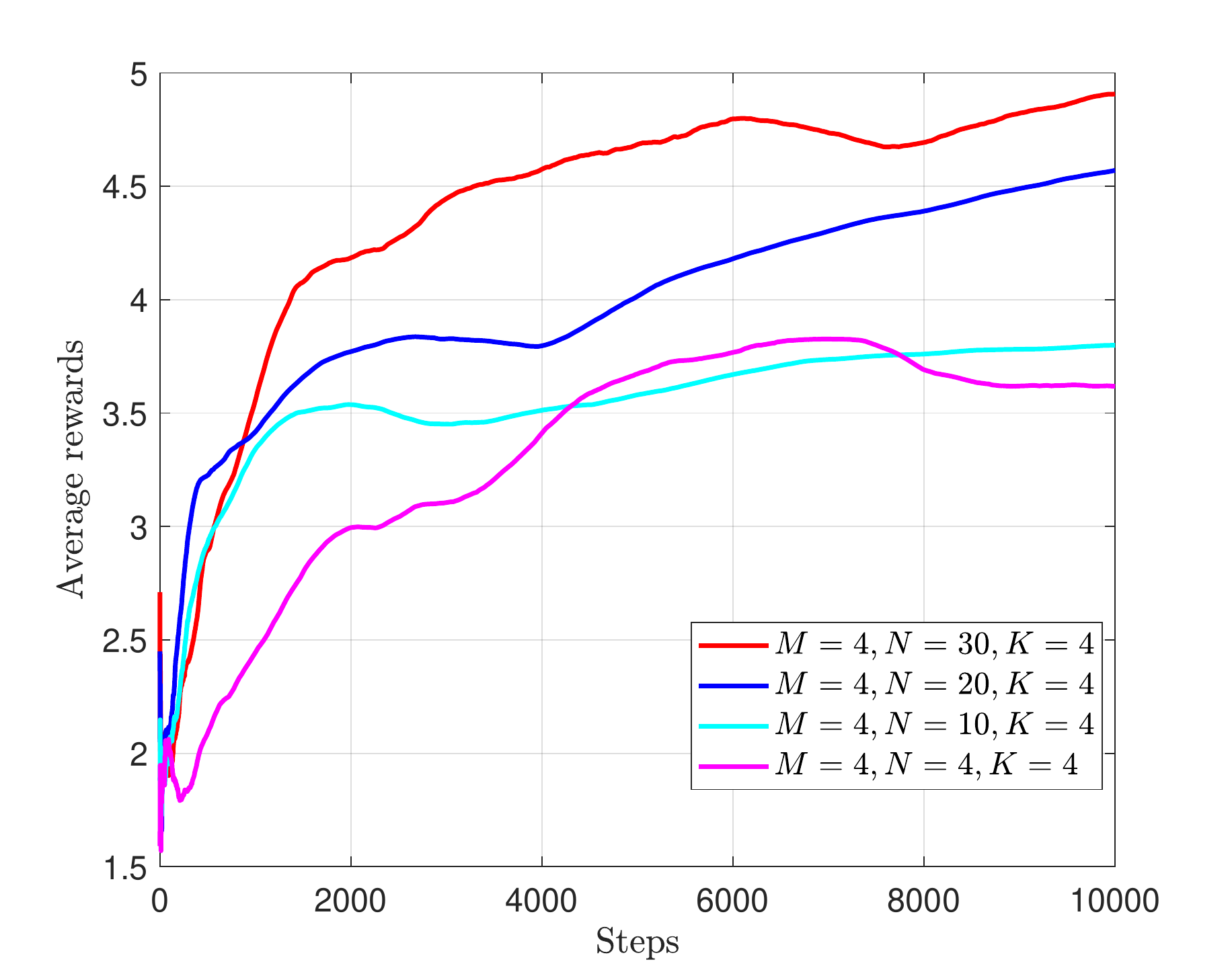}
  \caption{Average rewards versus time steps under different system parameter settings }.
  \label{fig:rewardn}
\end{center} \vspace{-6mm}
\end{figure}
\begin{figure}[htbp]
\begin{center}
  \includegraphics[width=9.3cm]{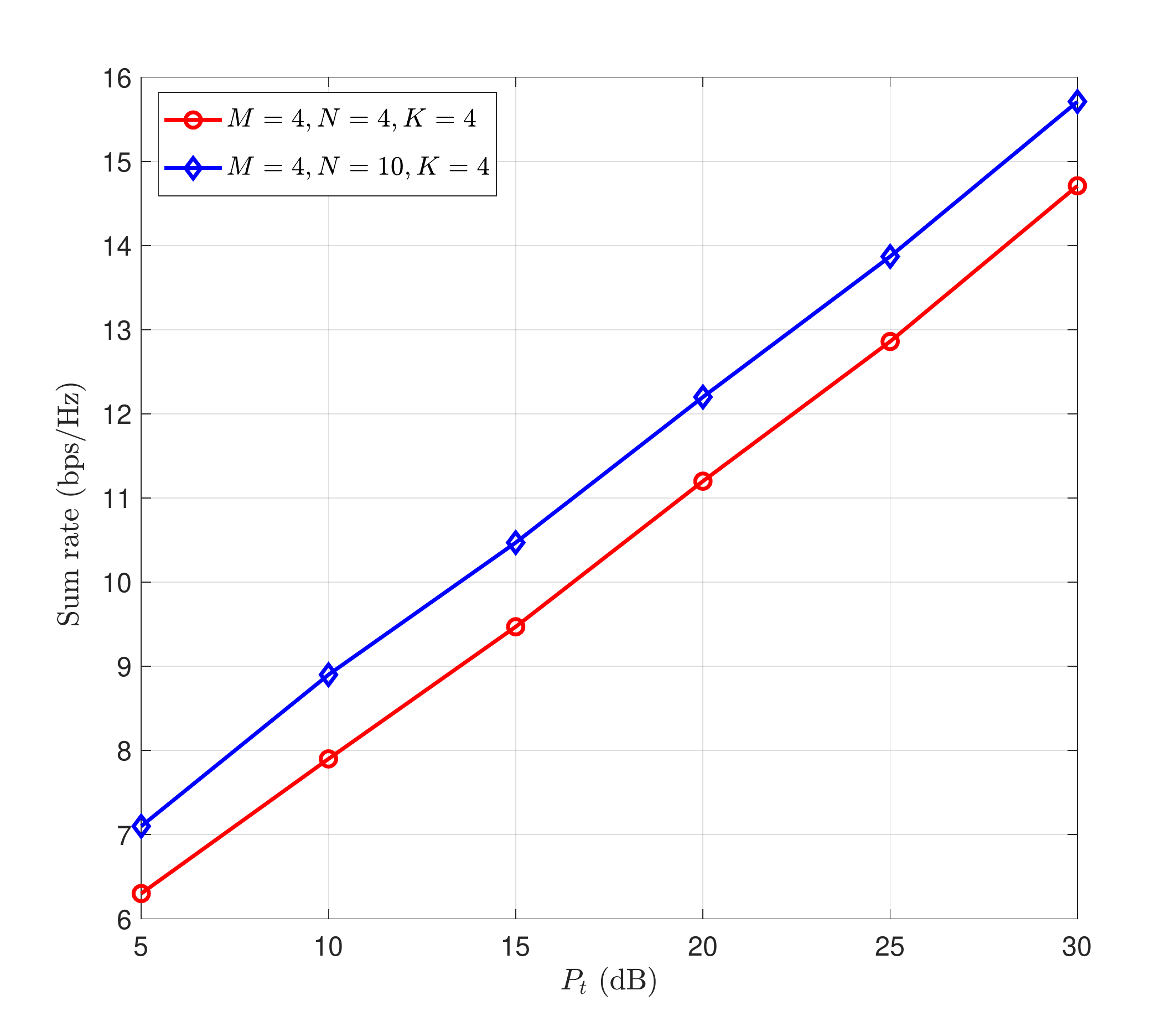}
  \caption{Sum rate as a function of $P_t$ under two scenarios}
  \label{fig:sumrate}
\end{center} \vspace{-6mm}
\end{figure}
It can be seen that, the rewards will converge with the increase of time step $t$. It converges faster at the low SNR ($P_t=5dB$) than high SNR ($P_t=20dB$). The reason is that, with higher SNR, the dynamic range of instant rewards is large, resulting in more fluctuations and worse convergence. These two figures also show that, starting from the identity matrices, the DRL based algorithm is able to learn from the environment and adjust $\mathbf{G}$ and $\mathbf{\Phi}$ to approach optimal solutions. Furthermore, the result of average rewards versus time steps under different $P_r=\{-10dB, 0dB, 10dB, 20dB, 30dB\}$ is shown in Fig. \ref{fig:rewardpt}. It can be seen that the SNRs have significantly effect on the convergence rate and performance, especially for the low SNR scenarios, i.e., below $10dB$.  When $P_t \geq 10dB$, the performance gap is far less than that between $P_t=0dB$ and $P_t=10dB$. In other words, the proposed DRL method is extremely sensitive to the low SNR although it takes less time to achieve the convergence.
\subsection{Impact of system settings}

Similarly, we investigate the impact of element number $N$ on the performance of DRL shown in Fig. \ref{fig:rewardn}, in which we considered the system settings $N=\{4,10,20,30\}$ with rewards versus time steps. Compared with the transmit power, DRL is more robust to the change of system settings. Specifically, with the increase of elements $N$, the average rewards also increase gradually as expected, but this doesn't increase the convergence time of the DRL method.

Fig. \ref{fig:sumrate} presents the average sum rate as a function of $P_t$. From this figure, we see that, the average sum rate increases with $P_t$. As more transmit power is allocated to the BS, higher average sum rate can be achieved by the proposed DRL based algorithm. This observation is aligned to that of conventional multiuser MISO systems. With joint design of transmit beamforming and phase shifts, the co-channel interference of multiuser MISO systems can be efficiently reduced, resulting in the performance improvement with $P_t$.

\begin{figure}[ht]
\begin{center}
\includegraphics[width=9.0cm]{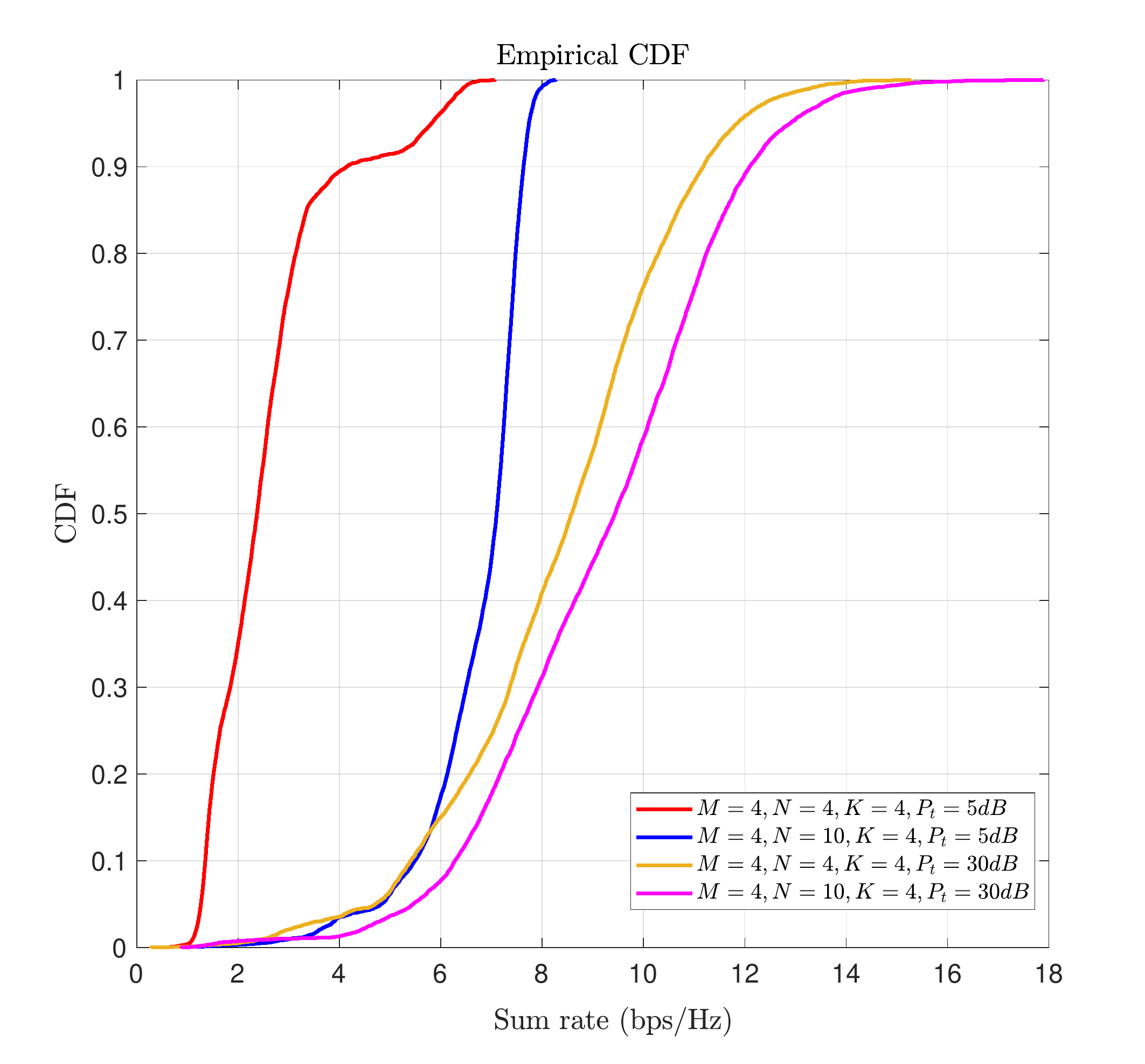}
\caption{CDF of sum rate for various system settings}
\label{fig:CDF} \vspace{-6mm}
\end{center}
\end{figure}

In Fig. \ref{fig:CDF}, we plot the cumulative distribution function (CDF) of the sum rate over different snapshots for different system settings. It is seen that the CDF curves confirm the observations from Fig. \ref{fig:sumrate}, where the average sum rates improve with the transmission power $P_t$ and the number of RIS elements $N$.

\subsection{Impact of learning and decaying rate}

\begin{figure}[t]
\begin{center}
  \includegraphics[width=9.5cm]{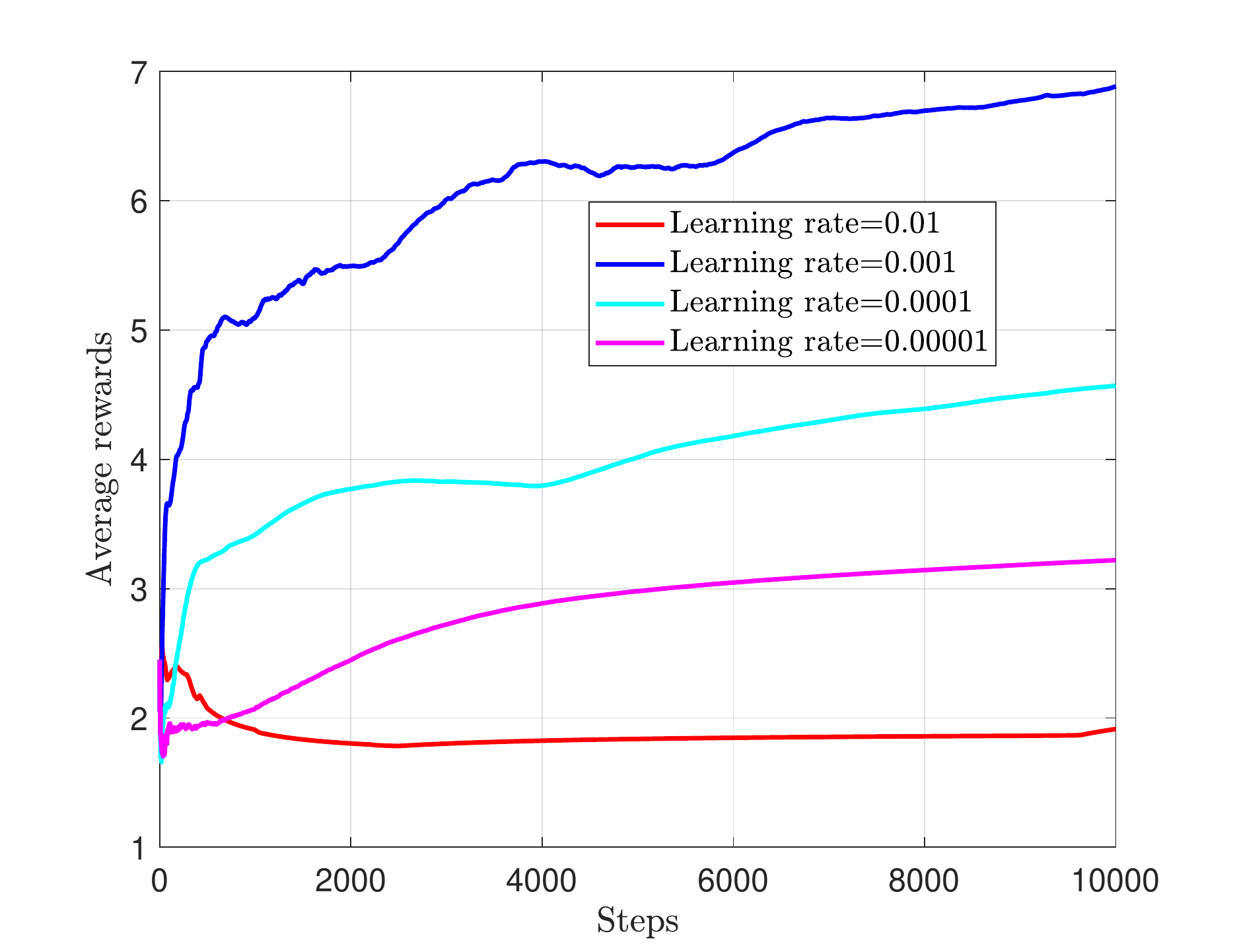}
  \caption{Average rewards versus steps under different learning rates, i.e., \{0.01, 0.001, 0.0001, 0.0001\}.}
  \label{fig:rewardlr}
\end{center} \vspace{-6mm}
\end{figure}

\begin{figure}[ht]
\begin{center}
  \includegraphics[width=9.5cm]{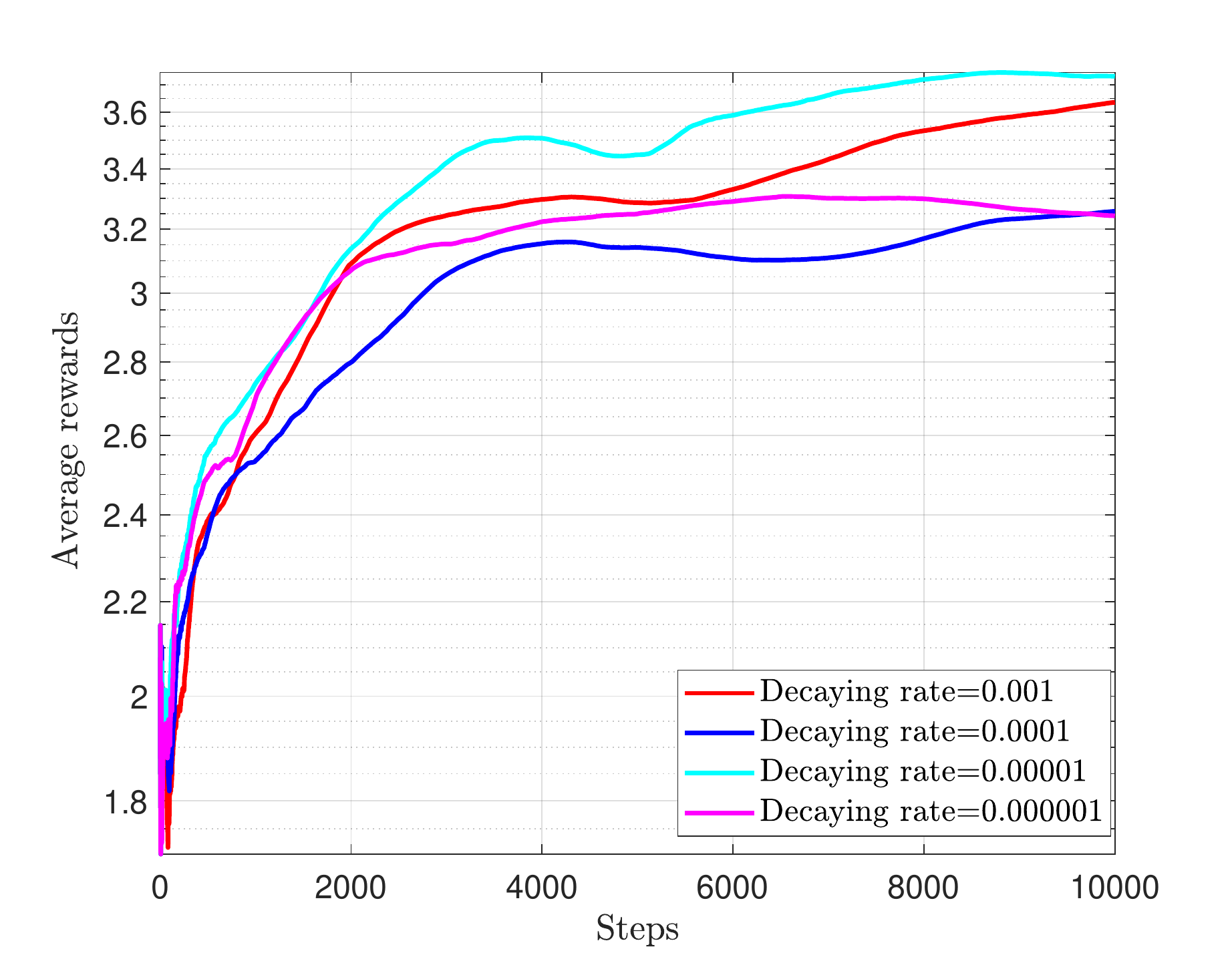}
  \caption{Average rewards versus steps under different decaying rates, i.e., \{0.001, 0.0001, 0.00001, 0.00001\}.}
  \label{fig:rewarddr}
\end{center} \vspace{-6mm}
\end{figure}
In our proposed DRL algorithm, we use constant learning and decaying rates for the critic and actor neural networks, and investigate their impacts on the performance and converge rate of DRL-based method. Fig. \ref{fig:rewardlr} demonstrates average rewards versus time steps under different learning rates, i.e., \{0.01, 0.001, 0.0001, 0.0001\}. It can be seen that different learning rates have  the great influence on the performance of the DRL algorithm. Specifically, the DRL with 0.001 learning rate achieves the best performance although it  takes a longer time to converge compared with 0.0001 and 0.00001 learning rate, while the large learning rate as 0.01 has the worse performance. This is because that too large learning rate will increases the oscillation that renders the performance drop dramatically. To sum up, the learning rate should be selected properly, neither too large nor too small. Fig. \ref{fig:rewarddr} compares average rewards versus time steps under different decaying rates, i.e., \{0.001, 0.0001, 0.00001, 0.00001\}. It shares the similar conclusion with the learning rate, but it exerts less influence on the DRL's performance and convergence rate. It can be seen that although 0.00001 decaying rate achieves the best performance, the gap between them are narrowed significantly.

Finally, we also should point out that, the performance of DRL based algorithms is very sensitive to initialization of the DNN and the other hyper-parameters, i.e., minibatch size, etc. The hyper-parameters need to be defined delicately under a given system setting, and the appropriate neural network hyper-parameters setting will improves significantly the performance of the proposed DRL algorithm as well as its convergence rate.

\section{Conclusions}
In this paper, a new joint design of transmit beamforming and phase shifts based on the recent advances in DRL technique was proposed, which attempts to formulate a framework that incorporates the DRL technique into the optimal designs for reflecting RIS assisted MIMO systems to address large-dimension optimization problems. The proposed DRL based algorithm has a very standard formulation and low complexity in implementation, without the knowledge of explicit mathematical formulations of wireless systems. It is therefore very easy to be scaled to accommodate various system settings. Moreover, the proposed DRL based algorithm is able to learn the knowledge about the environment and also is robust to the environment, through trial-and-error interactions with the environment by observing predefined rewards. Unlike most reported works utilizing alternating optimization techniques to alternatively obtain the transmit beamforming and phase shifts, the proposed DRL based algorithm obtains the joint design simultaneously as the output of the DNNs.  Simulation results show that the proposed DRL algorithm is able to learn from the environment through observing the instant rewards and improve its behavior step by step to obtain the optimal transmit beamforming matrix and phase shifts. It is also observed that, appropriate neural network parameter settings will improve significantly the performance and convergence rate of the proposed algorithm.

\end{document}